\documentclass[twocolumn,secnumarabic,amssymb, nobibnotes, aps, prd,english]{revtex4-1}
\usepackage[T1]{fontenc}
\usepackage[latin9]{inputenc}
\usepackage{amsmath}
\usepackage{amssymb}
\usepackage{graphicx}

\makeatletter
\@ifundefined{textcolor}{}
{%
 \definecolor{BLACK}{gray}{0}
 \definecolor{WHITE}{gray}{1}
 \definecolor{RED}{rgb}{1,0,0}
 \definecolor{GREEN}{rgb}{0,1,0}
 \definecolor{BLUE}{rgb}{0,0,1}
 \definecolor{CYAN}{cmyk}{1,0,0,0}
 \definecolor{MAGENTA}{cmyk}{0,1,0,0}
 \definecolor{YELLOW}{cmyk}{0,0,1,0}
}

\usepackage{babel}

\makeatother

\usepackage{babel}
\begin{document}

\title{CCCP Algorithms to Minimize the Bethe free energy of 3-SAT Problem }

\author{Yusupjan Habibulla$^{1}$}

\email{yusupjan29@itp.ac.cn}

\selectlanguage{english}%

\affiliation{$^{1}$State Key Laboratory of Theoretical Physics, Institute of
Theoretical Physics, Chinese Academy of Science, Beijing 100190, People's
Republic of China}

\newcommand{\revtex}{REV\TeX\ }
\newcommand{\classoption}[1]{\texttt{#1}}
\newcommand{\macro}[1]{\texttt{\textbackslash#1}}
\newcommand{\m}[1]{\macro{#1}}
\newcommand{\env}[1]{\texttt{#1}}
\setlength{\textheight}{9.5in}
\begin{abstract}
The k-sat problem is a prototypical constraint satisfaction problem.
There are many algorithms to study k-sat problem, BP algorithm is
famous one of them. But BP algorithm does not converge when $\alpha$(constraint
density)is bigger than some threshold value. In this paper we use
CCCP (Concave Convex Procedure) algorithm to study 3-sat problem and
we get better results than BP algorithm that CCCP algorithm still
converges when BP algorithm does not converge. Our work almost builds
on recent results by Yuille~\cite{Yuille2002} who apply the CCCP algorithm
to Bethe and Kikuchi free energies and obtained two algorithms on
2D and 3D spin glasses. Our implementation of CCCP algorithm on 3-sat
problem is some different from his implementation and we have some
different views about CCCP algorithm's some properties. Some difference
of these maybe because of CCCP algorithm have different properties
and implementation process on different problem and some others of
these are related to the CCCP algorithm itself. Our work indicates
that CCCP algorithm has more learning and inference applications.
\\

\end{abstract}

\date{\today}

\maketitle
\tableofcontents
\section{Introduction}

The k-sat problem is a prototypical constraint satisfaction problem
in the nondeterministic polynomial complete (NP-complete) complexity
class. There are exactly highly efficient algorithm for 2-sat problem
but perhaps there is no efficient algorithm for k-sat~\cite{Hartmann2005,Mezard2009}
(k\textgreater{}2) problem. The k-sat problem was intensively studied
in the statistical physics community during the last decade~\cite{Monasson1996,Krzakala2007,Domb1972}.
The BP (Belief Propagation) algorithm which comes from using variational
method to the statistical physics is very successful on k-sat problem.
The algorithm gives highly accurate results, but the algorithm does
not always converge, this is almost because of their error mode. This
paper using CCCP algorithm to study 3-sat problem and is guaranteed
to converge to the extrema of the Bethe free energy.

Belief Propagation\cite{Pearls1988} is a very powerful sum-product algorithm.
It is developed by Gallager for decoding Low Density Parity Codes
(LDPC)\cite{Gallager1963}. BP algorithm has been proven to converge for tree
like graphical models(Pearls 1988), but it always has been amazingly
successful when applied to inference problem with loops\cite{Freeman1999,Frey1998,Murphy1999},
and BP converges to almost good approximation to the true value of
Bethe approximation.

Statistical physics has been a powerful and rich idea resources for
statistical inference. The mean field approximation which can be represented
as minimizing a mean field free energy has been used to be resources
of optimization ideas\cite{A.LYuille1995}. The Bethe and Kikuchi free energiess
(Domb and Green 1972)contain higher order terms than the (factorized)
mean field free energies. So the Bethe and Kikuchi approximation give
more good results than standard mean field theory and is useful for
optimization and learning applications. There is a hierarchy of variational
approximation in statistical physics which starts with mean field,
proceeds to Bethe, and continuous with Kikuchi.

Yedidia et al's result\cite{Yedidia2000} proved that the fixed points of
BP correspond to the extrema points of the Bethe free energy functions.
They also developed a generalized belief propagation (GBP) algorithm
whose fixed points correspond to the extrema points of the Kikuchi
free energy. In practice, when BP and GBP converge they go to low
energy minima of the Bethe/Kikuchi free energies. In general, GBP
gives more accurate results than BP since Kikuchi is a better approximation
than Bethe. For example, empirically, GBP converged more close to
the true solution (Yedidia et al 2000) than BP algorithm on 2D spin
glasses. We can get the SGBP\cite{Wang.c2013,zhou2012}algorithm by simplifying
the GBP algorithm, and which convergence very quickly does than GBP.

The BP algorithm does not always converge, it is not convergent when
constraint density bigger than some threshold value for the k-sat
problem. So we search for other algorithms to minimize the Bethe/Kikuchi
free energies, CCCP algorithm is one of them.

The research on the belief propagation algorithm still continues untill
to now\cite{Zhou2013,Zhou2011,Shingh2013,Ahamadi2013}, and we can get more precise results
and more information about the system that we considered using
by RSB theory. There are a lot of work\cite{Gribova2002,Zeng2013,Likang2009,Xiao2011,Crisanti2002,Annibale2003} done
in this field.

It's main idea of the algorithms of developed using a Concave Convex
Procedure (CCCP) is decomposing the free energy into concave and convex
parts, and then construct discrete iterative rules which guareented
to decrease the free energy monotically. This procedure builds on
results developed when studying mean field theory\cite{J.Jkos1994,Ranga1996,Yuille1994}.

Yedidia (Yedidia et al 2000) using concave convex procedure (CCCP)
principle to get extrema point of Bethe and Kikuchi free energies,
this algorithm guaranteed to converge to extrema point of the Bethe
and Kikuchi free energy. Although this algorithm some like the BP/GBP
algorithm which estimate ``beliefs'' by propagating ``messages'',
but CCCP algorithm propagate messages (lagrange multipliers) by current
estimates of beliefs which must be re-estimated periodically.

This algorithm (CCCP algorithm) starts by decomposing the free energy
into concave and convex parts. From this decomposition we can get
discrete update rules which decrease the free energy at each iteration
step (but the free energy and entropy values are not very meaningful
in first several steps when constraint density larger than one it
is because of we have lagrange terms to add to convex free energy
parts which effects the free energies property, except all the constraint
and normalization conditions be satisfied). But the free energy is
meaningful at each iteration steps if we initialize the messages and
lagrange multipliers appropriately.

Yuille tested the algorithm on the 2D and 3D spin glasses in regular
graph. in his way this algorithm guaranteed to monotonically decrease
the free energy and the difference of beliefs. They randomly initialize
the vertical and horizontal potentials from gaussian distribution,
and randomly initialize the lagrange multipliers.

We tested CCCP principle on the 3-sat problem in random regular graph.
In this case in order to converge we must add another lagrange multiplier
(marginal normalization multiplier) to the lagrange parts (A.L.Yuille
didn't). This algorithm converges very rapidly if we implement it
parallel update way which gives incorrect results, and it doesn't
converge very quickly if we implement it in unparallel update way
which gives correct results. In our case this algorithm guaranteed
to minimize the free energy and difference of beliefs monotonically
at each iteration steps. Our results are very similar to the BP algorithm
when BP algorithm can converged. This algorithm still converge when
BP algorithm can't converge, but this gives strange (not meaningful)
results when $\alpha$ is(constraint density) bigger than some threshold
value (this value depend on the converge precision and the size of
system).

The structure of this paper is as follows. Section (\uppercase\expandafter{\romannumeral2}) describes 3-sat
problem and Bethe free energy, in section (\uppercase\expandafter{\romannumeral3}) we introduce the BP
equation. Section(\uppercase\expandafter{\romannumeral4}) we introduce CCCP principles and
we apply CCCP algorithm to 3-sat problem to determine the final form
of this double loop equation. Section(\uppercase\expandafter{\romannumeral5}) implementation CCCP algorithm
and compare this algorithm with BP algorithm, in section(\uppercase\expandafter{\romannumeral6}) we will
discuss some properties of CCCP algorithm, in the last section (\uppercase\expandafter{\romannumeral7})we
give our conclusion.

\section{Bethe free energy and 3-sat problem }

\subsection{Bethe free energy}

The Bethe free energy (Domb and Green 1972, Yedidia et al 2000) is
a variational technique from statistical physics. It's main idea is
to replace an hard to calculating problem that we can't calculate
by an approximation which is solvable. This approximation is using
joint distributions between variables which interacting each other.
It can give good results when standard mean field theory only gives
poor results. Consider a graph with nodes $i$=1,...,N. different
problem will determine different connections mode between the nodes.
If we just consider about two body interaction situations, then We
will only list connections between $ij$ for node pairs $i$, $j$
which are connected. The state of node is denoted by $x_{i}$. So
we can write the joint probability distribution function as
\begin{equation}
P(x_{1},...,x_{N})=\frac{1}{Z}\prod_{i,j:i>j}\psi_{ij}(x_{i},x_{j})\prod_{i}\psi_{i}(x_{i})
\end{equation}
where $\psi_{i}(x_{i})$=$e^{-{\beta}E_{i}(x_{i})}$ is the factor
of the node $i$, $Z$ is a normalization constant, and $\psi_{ij}(x_{i},x_{j})$=$e^{-{\beta}E_{ij}(x_{i},x_{j})}$
is the factor of the interaction between nodes $i$ and $j$. We use
the convention $i>j$ to avoid double counting (Attention: if node
$i$ and node $j$ not connected, then we do not have the term $\psi_{ij}$).
And then we can write the Bethe free energy as(Yedidia et al 2000)

\begin{flalign}
 & \begin{alignedat}{1}F_{\beta}(\{b_{ij},b_{i}\}) & =\!\!\!\sum_{i,j:i>j}\sum_{x_{i},x_{j}}b_{ij}(x_{i},x_{j})\log\frac{b_{ij}(x_{i},x_{j})}{\phi_{ij}(x_{i},x_{j})}\\
 & -\sum_{i}(n_{i}-1)\sum_{x_{i}}b_{i}(x_{i})\log\frac{b_{i}(x_{i})}{\psi_{i}(x_{i})}
\end{alignedat}
\end{flalign}
 where $\phi_{ij}(x_{i},x_{j})$=$\psi_{i}(x_{i})\psi_{ij}(x_{i},x_{j})\psi_{j}(x_{j})$
,$n_{i}$ is connectivity of the nodes $i$ or variable degree of
node $i$ in other words, the marginal probability \{$b_{i}$($x_{i}$)\},
and joint probability \{$b_{ij}(x_{i},x_{j})$\} must satisfy the
linear consistency constraints.

\begin{equation}
\sum_{x_{i},x_{j}}b_{ij}(x_{i},x_{j})=1,\forall i,j:i>j\sum_{x_{i}}b_{i}(x_{i})=1,\forall i,
\end{equation}

\begin{equation}
\sum_{x_{i}}b_{ij}(x_{i},x_{j})=b_{j}(x_{j}),\forall j,x_{j},\sum_{x_{j}}b_{ij}(x_{i},x_{j})=b(x_{i}),\forall i,x_{i}.
\end{equation}

So we can write the lagrange parts of the free energy as

\begin{equation}
\begin{aligned}\sum_{ij:i>j}\gamma_{ij}\{\sum_{x_{i},x_{j}}b_{ij}(x_{i},x_{j})-1\}\\
+\sum_{i,j:i>j}\sum_{x_{j}}\lambda_{ij\rightarrow{j}}(x_{j})\{\sum_{x_{i}}b_{ij}(x_{i},x_{j})-b_{j}(x_{j})\}\\
+\sum_{i,j:i>j}\sum_{x_{i}}\lambda_{ij\rightarrow{i}}(x_{i})\{\sum_{x_{j}}b_{ij}(x_{i},x_{j})-b_{i}(x_{i})\}
\end{aligned}
\end{equation}

If we consider many body interaction system, in the same way we can
write the joint probability distribution function as

\begin{equation}
P(x_{1},...,x_{N})=\frac{1}{Z}\prod_{\alpha}\psi_{\alpha}(x_{\alpha})\prod_{i}\psi_{i}(x_{i})
\end{equation}
where The state of node $i$ is denoted by $x_{i}$, and the state
of interaction $\alpha$ is denoted by $x_{\alpha}$. for example
if we consider about the $n$ body interaction system, the interaction
has $k^{n}$ ($k$ is the possible state number of one variable node)
possible states. $\psi_{i}(x_{i})$=$e^{-{\beta}E_{i}(x_{i})}$ is
the factor of the node $i$, $Z$ is a normalization constant, and
$\psi_{\alpha}(x_{\alpha})$=$e^{-{\beta}E_{\alpha}(x_{\alpha})}$
is the factor of the interaction $\alpha$. And then we can write
the Bethe free energy as(Yedidia et al 2000)

\begin{equation}
\begin{aligned}F_{\beta}(\{b_{\alpha},b_{i}\}) & =\sum_{\alpha}\sum_{x_{\alpha}}b_{\alpha}(x_{\alpha})\log\frac{b_{\alpha}(x_{\alpha})}{\phi_{\alpha}(x_{\alpha})}\\
 & -\sum_{i}(n_{i}-1)\sum_{x_{i}}b_{i}(x_{i})\log\frac{b_{i}(x_{i})}{\psi_{i}(x_{i})}
\end{aligned}
\end{equation}
where $\phi_{\alpha}(x_{\alpha})$=$\psi_{\alpha}(x_{\alpha})\prod_{i\in\alpha}\psi_{i}(x_{i})$
,$n_{i}$ is the variable degree of
node $i$, the marginal probability \{$b_{i}$($x_{i}$)\},
and joint probability \{$b_{\alpha}(x_{\alpha})$\} must satisfy the
linear consistency constraints.

\begin{equation}
\sum_{x_{\alpha}}b_{\alpha}(x_{\alpha})=1,\forall\alpha,\sum_{x_{i}}b_{i}(x_{i})=1,\forall i,
\end{equation}

\begin{equation}
\sum_{x_{\alpha}\setminus{x_{i}}}b_{\alpha}(x_{\alpha})=b_{i}(x_{i}),\forall i\in\alpha
\end{equation}

So we can write the lagrange parts of the free energy as

\begin{equation}
\begin{aligned} & \sum_{\alpha}\gamma_{\alpha}\{\sum_{x_{\alpha}}b_{\alpha}(x_{\alpha})-1\}\\
 & +\sum_{\alpha}\sum_{i\in\alpha}\sum_{x_{i}}\lambda_{\alpha\rightarrow{i}}(x_{i})
 \{\sum_{x_{\alpha}\setminus{x_{i}}}b_{\alpha}(x_{\alpha})-b_{i}(x_{i})\}
\end{aligned}
\end{equation}

\subsection{3-sat problem}

A 3-sat formula contain N variable nodes and M constraint nodes. Every
variable nodes only has two states \{+1 , -1\}, every constraint nodes
connect with 3 variable nodes and every connection represent one requirements
(request the variable must be +1, or -1)to the connected variables,
the constraint be satisfied when at least one of these three requirements
be satisfied. So we can write the energy function of the 3-sat problem
as

\begin{equation}
E=\sum_{\alpha=1}^{M}\prod_{i\in\Game\alpha}(\frac{1-J_{\alpha}^{i}x_{i}}{2})
\end{equation}

where the $J_{\alpha}^{i}$ is the requirement of the constraint node
$\alpha$ to the variable node $i$, if at least one variable of three
variables which connected by one constraint node satisfy the requirement,
 then the energy of the constraint node equal to zero. Otherwise
 the energy of the constraint node equal to one.

So we can write the distribution of one constraint nodes as

\begin{equation}
\begin{aligned}\psi_{ijk}(x_{i},x_{j},x_{k}) & =e^{-\beta E_{\alpha}}=e^{-\beta\prod_{i\in\Game\alpha}(\frac{1-J_{\alpha}^{i}x_{i}}{2})}\\
 & =e^{-\beta(\frac{1-J_{\alpha}^{i}x_{i}}{2})(\frac{1-J_{\alpha}^{j}x_{j}}{2})(\frac{1-J_{\alpha}^{k}x_{k}}{2})}
\end{aligned}
\end{equation}

Because of there is no external field so
\begin{equation}
E_{i}=0,\psi_{i}(x_{i})=1
\end{equation}

\begin{equation}
\phi_{ijk}=\psi_{i}\psi_{j}\psi_{k}\psi_{ijk}=\psi_{ijk}
\end{equation}

According to the equation (7) we can write the Bethe free energy for 3-sat problem as

\begin{equation}
\begin{aligned}\begin{aligned}F_{\beta}(\{b_{ijk},b_{i}\}) & =\sum_{\alpha}\sum_{x_{i},x_{j},x_{k}}b_{ijk}(x_{i},x_{j},x_{k})
\log\frac{b_{ijk}(x_{i},x_{j},x_{k})}{\phi_{ijk}(x_{i},x_{j},x_{k})}\\
 & -\sum_{i}(n_{i}-1)\sum_{x_{i}}b_{i}(x_{i})\log\frac{b_{i}(x_{i})}{\phi_{i}(x_{i})}
\end{aligned}
\end{aligned}
\end{equation}

And according to the equation (10), the lagrange multipliers part can be written as

\begin{equation}
\begin{aligned}\sum_{ijk}\gamma_{ijk}\{\sum_{x_{i},x_{j},x_{k}}b_{ijk}(x_{i},x_{j},x_{k})-1\}\\
+\sum_{i,j,k}\sum_{x_{k}}\lambda_{ijk\rightarrow{k}}(x_{k})\{\sum_{x_{i},x_{j}}b_{ijk}(x_{i},x_{j},x_{k})-b_{k}(x_{k})\}\\
+\sum_{i,j,k}\sum_{x_{j}}\lambda_{ijk\rightarrow{j}}(x_{j})\{\sum_{x_{k},x_{i}}b_{ijk}(x_{i},x_{j},x_{k})-b_{j}(x_{j})\}\\
+\sum_{i,j,k}\sum_{x_{i}}\lambda_{ijk\rightarrow{i}}(x_{i})\{\sum_{x_{j},x_{k}}b_{ijk}(x_{i},x_{j},x_{k})-b_{i}(x_{i})\}
\end{aligned}
\end{equation}

\section{BP equation}

Directly compute the partition function is a hard task in large system,
but this complexity can be reduced when the underlying factor graph
has some special structure. In factor graph N variables can be expressed
by N variable nodes (circle empty nodes), M interactions can be expressed
by function nodes (square filled nodes). So we can write the partition
function as

\begin{equation}
Z=\sum_{\underline{\sigma}}\prod_{i=1}^{N}\psi_{i}(\sigma_{i})\sum_{\underline{\sigma}_{\partial a}}\prod_{a=1}^{M}\psi_{a}(\underline{\sigma}_{\partial a})\prod_{(j,a)\in G}\delta(\sigma_{j}^{a},\sigma_{j})
\end{equation}

where the $\underline{\sigma}$ represents all the configurations
of the $N$ variables system, $\underline{\sigma}_{\partial a}$ represents
all the configurations of the interaction $a$, $\psi_{i}(x_{i})$=$e^{-{\beta}E_{i}({\sigma}_{i})}$
represents the external field factor of the variable node $i$,
 $\psi_{a}(\underline{\sigma}_{\partial a})$=$e^{{-\beta}E_{a}(\underline{\sigma}_{\partial a})}$
represents the interaction field factor of the clause node $a$.

Now we introduce the edge Auxiliary probability function to the partition
function, so

\begin{equation}
\begin{aligned}Z & =\sum_{\underline{\sigma}}\prod_{i=1}^{N}\psi_{i}(\sigma_{i})\sum_{\underline{\sigma}_{\partial a}}\prod_{a=1}^{M}[\psi_{a}(\underline{\sigma}_{\partial a})\prod_{j\in\partial a}q_{j\rightarrow a}(\sigma_{j}^{a})]\\
 & \times\prod_{(k,b)\in G}\frac{\delta(\sigma_{k}^{b},\sigma_{k})}{q_{k\rightarrow b}(\sigma_{k}^{b})}
\end{aligned}
\end{equation}

in there, the edge Auxiliary probability function satisfy the normalization
condition.

And we introduce the another normalized edge Auxiliary probability
function to the partition function, so

\begin{equation}
\begin{aligned}Z & =\sum_{\underline{\sigma}}\prod_{i=1}^{N}\psi_{i}(\sigma_{i})\prod_{a\in\partial i}p_{a\rightarrow i}(\sigma_{i})\times\\
&\sum_{\underline{\sigma}_{\partial b}}
\prod_{b=1}^{M}[\psi_{b}(\underline{\sigma}_{\partial b})\prod_{j\in\partial b}q_{j\rightarrow b}(\sigma_{j}^{b})]
  \prod_{(k,c)\in G}\frac{\delta(\sigma_{k}^{c},\sigma_{k})}{q_{k\rightarrow c}(\sigma_{k}^{c})p_{c\rightarrow k}(\sigma_{k}^{c})}
\end{aligned}
\end{equation}

Eventually we can write the partition function as

\begin{equation}
Z=Z_{0}(1+\sum_{g\in G}L_{g})
\end{equation}

in there $L_{g}$ represents the loopy factor graph distribution of
our system to the partition function, and
\begin{equation}
Z_{0}=\frac{\prod_{i\in G}z_{i}\prod_{a\in G}z_{a}}{\prod_{(i,a)\in G}z_{(i,a)}}
\end{equation}

\begin{equation}
z_{i}=\sum_{\sigma_{i}}\psi_{i}(\sigma_{i})\prod_{a\in\partial i}p_{a\rightarrow i}(\sigma_{i})
\end{equation}

\begin{equation}
z_{a}=\sum_{\underline{\sigma}_{\partial a}}\psi_{a}(\underline{\sigma}_{\partial a})\prod_{i\in\partial a}q_{i\rightarrow a}(\sigma_{i}^{a})
\end{equation}

\begin{equation}
z_{(i,a)}=\sum_{\sigma_{i}}p_{a\rightarrow i}(\sigma_{i})q_{i\rightarrow a}(\sigma_{i})
\end{equation}

If these distributions of the loopy factor graphs (Marc Mezard, Andrea
Montanari. 2009) are equal to zero, then we can write the partition
function as

\begin{equation}
Z=Z_{0}
\end{equation}

In this case we can easily to calculate all the thermodynamical functions.

If and only if the edge Auxiliary probability function satisfy the
following iterative equations, and then these distributions of the
loopy factor graphs are equal to zero.

\begin{equation}
q_{i\rightarrow a}(\sigma)=\frac{1}{z_{i\rightarrow a}}\psi_{i}(\sigma)\prod_{b\in\partial i\setminus a}p_{b\rightarrow i}(\sigma)
\end{equation}

\begin{equation}
p_{a\rightarrow i}(\sigma)=\frac{1}{z_{a\rightarrow i}}\sum_{\underline{\sigma}_{\partial a}}\delta(\sigma_{i},\sigma)\psi_{a}(\underline{\sigma}_{\partial a})\prod_{j\in\partial a\setminus i}q_{j\rightarrow a}(\sigma_{j})
\end{equation}

These equations called as Belief Propagation equations(abbreviation
BP equations).

In this equations

\begin{equation}
z_{i\rightarrow a}=\sum_{\sigma}\psi_{i}(\sigma)\prod_{b\in\partial i\setminus a}p_{b\rightarrow i}(\sigma)
\end{equation}

\begin{equation}
z_{a\rightarrow i}=\sum_{\underline{\sigma}_{\partial a}}\psi_{a}(\underline{\sigma}_{\partial a})\prod_{j\in\partial a\setminus i}q_{j\rightarrow a}(\sigma_{j})
\end{equation}

So if the factor graph has tree like structure (or no contain loopy
structure), we can calculate the partition function with very simplest
form by iterating the BP equations and getting the stable points of
Auxiliary probability functions. But in the 3-sat problem the BP equations
are not convergent when constraint density greater than 3.86 in big
system.

\section{Concave Convex Procedure(CCCP) algorithm }

The main idea of the CCCP algorithm is that first step decomposing
the free energy into two parts respectively convex part and concave part.
In the second step we add the constraint condition to the convex part.
Find the minimum point of the bethe free energy by dynamical programming
procedure. The algorithm iterates by matching points on the two curves(convex
and concave) that have the same tangent vectors.

\begin{figure}[h!]
\includegraphics[scale=0.35]{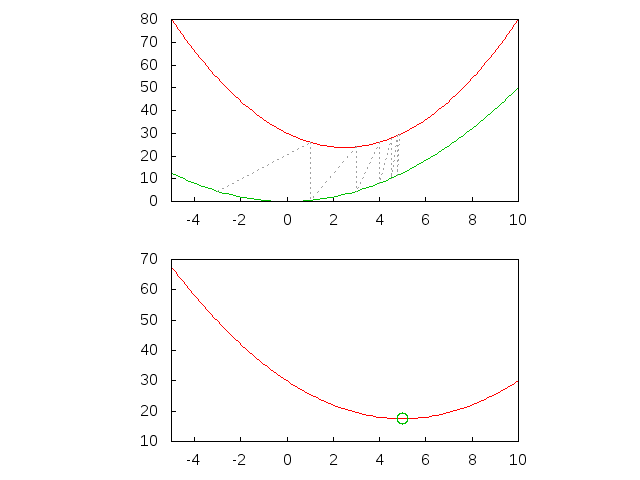}

\caption{A CCCP algorithm illustrated for convex minus convex. We want
to minimize the function in the bottom panel. We decompose it (top panel) into
a convex part (top curve) minus a convex term (bottom curve). The algorithm
iterates by matching points on the two curves that have the same tangent vectors.
See the text for more details. The algorithm rapidly converges to the solution at
x=5.0.}
\end{figure}

Below we simplest way to recall the process of the derivation of CCCP
algorithm:

\subsection{the minimum point of the Bethe free energy without constraints}

Theorem 1. Consider an energy function E($\vec{z}$) (bounded below)
of form E($\overrightarrow{z}$) = $E_{vex}$($\overrightarrow{z}$)
+ $E_{cave}$($\overrightarrow{z}$) where $E_{vex}$($\overrightarrow{z}$)
and $E_{cave}$($\overrightarrow{z}$) are convex and concave functions
of $\overrightarrow{z}$ respectively. Then the discrete iterative
algorithm $\vec{z}^{t}$$\longmapsto$ $\vec{z}^{t+1}$given by:
\[
\nabla E_{vex}(\vec{z}^{t+1})=-\nabla E_{cave}(\vec{z}^{t}),
\]

we can prove that this equation guaranteed to minimize the free energy.

Proof: corresponding to the property of concavity and convexity functions,
we can write following inequation

\begin{equation}
E_{vex}(\vec{z}_{2})\geqslant E_{vex}(\vec{z}_{1})+(\vec{z}_{2}-\vec{z}_{1})\cdot\nabla E_{vex}(\vec{z}_{1})
\end{equation}

\begin{equation}
E_{cave}(\vec{z}_{4})\leqslant E_{cave}(\vec{z}_{3})+(\vec{z}_{4}-\vec{z}_{3})\cdot\nabla E_{cave}(\vec{z}_{3}),
\end{equation}

now we set $\vec{z}_{1}=\vec{z}^{t+1}$,$\vec{z}_{2}=\vec{z}^{t}$,$\vec{z}_{3}=\vec{z}^{t}$,$\vec{z}_{4}=\vec{z}^{t+1}$using
upper three equation we obtain that:

\begin{equation}
E_{vex}(\vec{z}^{t+1})+E_{cave}(\vec{z}^{t+1})\leqslant E_{vex}(\vec{z}^{t})+E_{cave}(\vec{z}^{t}),
\end{equation}

which prove the claim.

\subsection{the minimum point of the Bethe free energy with constraints}

Theorem 2. Consider a function E($\overrightarrow{z}$) = $E_{vex}$($\overrightarrow{z}$)
+ $E_{cave}$($\overrightarrow{z}$) subject to k linear constraints
$\vec{\phi}^{\mu}.\overrightarrow{z}=c^{\mu}$where \{$c^{\mu}:\mu=1,....k$\}are
constants. Then the algorithm $\vec{z}^{t}\longmapsto\vec{z}^{t+1}$given
by
\begin{equation}
\nabla E_{vex}(\vec{z}^{t+1})=-\nabla E_{cave}(\vec{z}^{t})-\sum_{\mu=1}^{k}\alpha^{\mu}\vec{\phi}^{\mu},
\end{equation}

where the parameters \{$\alpha^{\mu}$\}are chosen to ensure that
$\vec{z}^{t+1}.\vec{\phi}^{\mu}=c^{\mu}$for $\mu=1,...k,$

Proof: first we define orthogonal unit vectors$\{\vec{\psi}^{\nu}:\nu=1,...n-k\}$,
which span the space orthogonal to the constraints \{$\vec{\phi}^{\mu}:\mu=1,...k$\}.
Let $\vec{y}(\vec{z})=\sum{\scriptscriptstyle _{_{\nu=1}}^{^{n-k}}}\vec{\psi}^{\nu}(\vec{z}.\vec{\psi}^{\nu})$
(project the z to different vectors or on the other hand express the
z in \{$\psi^{\nu}$\}space). Define function $\hat{E_{vex}}(\vec{y}),\hat{E_{cave}}(\vec{y})$,
which ensure that $\hat{E_{vex}}(\vec{y}(\vec{z}))=E_{vex}(\overrightarrow{z}),
\hat{E_{cave}}(\vec{y}(\vec{z}))=E_{cave}(\overrightarrow{z})$.

So we can write
\begin{equation}
\vec{\psi}^{\nu}.\vec{\nabla}_{\vec{z}}E_{vex}(\vec{z}^{t+1})=
\vec{-\psi}^{\nu}.\vec{\nabla}_{\vec{z}}E_{cave}(\vec{z}^{t})
\end{equation}

where $\nu=1,...n-k.(\vec{\psi}^{\nu}.\vec{\phi}^{\mu}=0)$.

It follows from theorem2 that
\begin{equation}
\bar{E}_{vex}(\vec{z}^{t+1})=E_{vex}(\vec{z}^{t+1})+\sum_{\mu}\alpha^{\mu}\{\vec{\phi}^{\mu}.\vec{z}^{t+1}-c^{\mu}\}
\end{equation}

So we need to impose the constraints only on convex terms.

\subsection{the solution of the cccp algorithm}

Theorem 3. let $E_{vex}^{t+1}(\vec{z})=\sum_{i}z_{i}\log\frac{z_{i}}{\xi_{i}}$.
Then the update equation of theorem 2 can be expressed as minimizing
the convex energy function:
\begin{equation}
\begin{aligned}E^{t+1}(\vec{z}^{t+1}) & =\vec{z}^{t+1}.\vec{h}+\sum_{i}z_{i}^{t+1}\log\frac{z_{i}^{t+1}}{\xi_{i}}\\
 & +\sum_{\mu}\alpha^{\mu}\{\vec{\phi}^{\mu}.\vec{z}^{t+1}-c^{\mu}\}
\end{aligned}
\end{equation}

Where $\vec{h}=\vec{\nabla}E_{cave}(\vec{z}^{t})$, and $\vec{z}^{t+1}.\vec{h}=\sum_{i}\vec{z}_{i}^{t+1}\vec{h}_{i}$The
solution of the form is
\begin{equation}
z_{i}^{t+1}(\alpha)=\xi_{i}e^{-h_{i}}e^{-1}e^{-\sum_{\mu}\alpha^{\mu}\phi^{\mu}}
\end{equation}

Proof: insert the convex free energy function to the theorem 2 equation,
then we can get
\begin{equation}
\log\frac{z_{i}^{t+1}}{\xi_{i}}+1+\sum_{\mu}\alpha^{\mu}\phi^{\mu}=-h_{i}
\end{equation}

Where $h_{i}=\vec{\nabla_{z_{i}}}E_{cave}(\vec{z}^{t})$

\begin{equation}
z_{i}^{t+1}(\alpha)=\xi_{i}e^{-h_{i}}e^{-1}e^{-\sum_{\mu}\alpha^{\mu}\phi_{i}^{\mu}}
\end{equation}

where the lagrange multipliers $\{\alpha^{\mu}\}$ are constrained
to maximize the dual energy

\begin{equation}
\hat{E}^{t+1}(\alpha)=-\sum_{i}z_{i}^{t+1}(\alpha)-\sum_{\mu}\alpha^{\mu}c^{\mu}
\end{equation}

\[
\hat{E}^{t+1}(\alpha)=-\sum_{i}\xi_{i}e^{-h_{i}}e^{-1}
e^{-\sum_{\mu}\alpha^{\mu}\phi_{i}^{\mu}}-\sum_{\mu}\alpha^{\mu}c^{\mu}
\]

Moreover, maximizing $\hat{E}^{t+1}(\alpha)$ with respect to specific
$\alpha^{\mu}$ enables us to satisfy the corresponding constraint
exactly.

\begin{widetext}

\subsection{convex and concave procedure (CCCP) algorithm for the 3-sat problem}

Theorem 3 specifies a double loop algorithm where the outer loop is
given by the solution equation and the inner loop equation is given
by the normalization condition and constraint condition, and the inner
loop determines the \{$\alpha^{\mu}$\} by maximizing the dual energy.

\begin{equation}
\begin{aligned}F_{vex}=\sum_{ijk}\sum_{x_{i},x_{j},x_{k}}b_{ijk}(x_{i},x_{j},x_{k})
\log\frac{b_{ijk}(x_{i},x_{j},x_{k})}{\psi_{ijk}(x_{i},x_{j},x_{k})}+
\sum_{i}\sum_{x_{i}}b_{i}(x_{i})\log\frac{b_{i}(x_{i})}{\psi_{i}(x_{i})}\end{aligned}
\end{equation}
\begin{equation}
F_{cave}=-\sum_{i}n_{i}\sum_{x_{i}}b_{i}(x_{i})\log\frac{b_{i}(x_{i})}{\psi_{i}(x_{i})}
\end{equation}

The outer loop equations

\begin{equation}
b_{ijk}(x_{i},x_{j},x_{k};t+1)=\psi_{ijk}(x_{i},x_{j},x_{k})e^{-1}
e^{-\lambda_{ijk\rightarrow{i}}(x_{i})}e^{-\lambda_{ijk\rightarrow{j}}(x_{j})}
e^{-\lambda_{ijk\rightarrow{k}}(x_{k})}e^{-\gamma_{ijk}}
\end{equation}

\begin{equation}
b_{i}(x_{i};t+1)=\psi_{i}(x_{i})e^{-1}e^{n_{i}}[b(x_{i};t)]^{n_{i}}
e^{\sum_{j}\sum_{k}\sum_{l}\lambda_{jkl\rightarrow{i}}(x_{i})}
\end{equation}

The inner loop equations

\begin{equation}
e^{\gamma_{ijk}(t+1)}=\sum_{x_{i},x_{j},x_{k}}\psi_{ijk}(x_{i},x_{j},x_{k})
e^{-1}e^{-\lambda_{ijk\rightarrow{i}}(x_{i};t)}e^{-\lambda_{ijk\rightarrow{j}}(x_{j};t)}
e^{-\lambda_{ijk\rightarrow{k}}(x_{k};t)}
\end{equation}
\begin{equation}
e^{2\lambda_{ijk\rightarrow{k}}(x_{k};t+1)}=\frac{\sum_{x_{i},x_{j}}\psi_{ijk}(x_{i},x_{j},x_{k})
e^{-\lambda_{ijk\rightarrow{i}}(x_{i};t)}e^{-\lambda_{ijk\rightarrow{j}}(x_{j};t)}
e^{-\gamma_{ijk}(t)}}{e^{n_{k}}(b_{k}(x_{k};t))^{n_{k}}e^{\sum_{o\neq i}{\sum_{p\neq j}\sum_{q\neq k}\lambda_{opq\rightarrow{k}}(x_{k};t)}}}
\end{equation}
\begin{equation}
e^{2\lambda_{ijk\rightarrow{j}}(x_{j};t+1)}=\frac{\sum_{x_{k},x_{i}}\psi_{ijk}(x_{i},x_{j},x_{k})
e^{-\lambda_{ijk\rightarrow{k}}(x_{k};t)}e^{-\lambda_{ijk\rightarrow{i}}(x_{i};t)}
e^{-\gamma_{ijk}(t)}}{e^{n_{j}}(b_{j}(x_{j};t))^{n_{j}}e^{\sum_{o\neq i}{\sum_{p\neq j}\sum_{q\neq k}\lambda_{opq\rightarrow{j}}(x_{j};t)}}}
\end{equation}
\begin{equation}
e^{2\lambda_{ijk\rightarrow{i}}(x_{i};t+1)}=\frac{\sum_{x_{j},x_{k}}\psi_{ijk}(x_{i},x_{j},x_{k})
e^{-\lambda_{ijk\rightarrow{j}}(x_{j};t)}e^{-\lambda_{ijk\rightarrow{k}}(x_{k};t)}e^{-\gamma_{ijk}(t)}}
{e^{n_{i}}(b_{i}(x_{i};t))^{n_{i}}e^{\sum_{o\neq i}{\sum_{p\neq j}\sum_{q\neq k}\lambda_{opq\rightarrow{i}}(x_{i};t)}}}
\end{equation}

BP style for outer loop

\begin{equation}
\begin{aligned}b_{ijk}(x_{i},x_{j},x_{k};t+1) & =\psi_{ijk}(x_{i},x_{j},x_{k})e^{-1}e^{-\lambda_{j\rightarrow\alpha}(x_{j})}
e^{-\lambda_{i\rightarrow\alpha}(x_{i})}e^{-\lambda_{k\rightarrow\alpha}(x_{k})}e^{-\gamma_{\alpha}}\\
 & =\psi_{ijk}(x_{i},x_{j},x_{k})e^{-1}\prod_{j\in\partial\alpha}
 e^{-\lambda_{j\rightarrow\alpha}(x_{j})}e^{-\gamma_{\alpha}}
\end{aligned}
\end{equation}

\begin{equation}
b_{i}(x_{i};t+1)=\psi_{i}(x_{i})e^{-1}e^{n_{i}}[b(x_{i};t)]^{n_{i}}e^{\sum_{\alpha\in\partial i}\lambda_{\alpha\rightarrow i}(x_{i})}=\psi_{i}(x_{i})e^{-1}e^{n_{i}}[b(x_{i};t)]^{n_{i}}\prod_{\alpha\in\partial i}e^{\lambda_{\alpha\rightarrow i}(x_{i})},
\end{equation}

where $\lambda_{j\rightarrow\alpha}(x_{j})=\lambda_{\alpha\rightarrow j}(x_{j})=\lambda_{ijk\rightarrow{j}}(x_{j})$
. We can see the CCCP algorithm updating data on  unidirectional graph, but the BP algorithm do on bidirectional graph.

\subsection{discussion and the final form of the double loop equation}

Because of these equations can not guaranteed to satisfy the constraint condition in simulation,
so the lagrange multipliers always converge to NAN, it leads to the
marginal beliefs converge to NAN. In order to avoid this we must normalize
the marginal beliefs by adding another lagrange multipliers.

\begin{equation}
\sum_{i}\alpha_{i}\{\sum_{x_{i}}b_{i}(x_{i})-1\}
\end{equation}

Our outer loop equation

\begin{alignat}{1}
\begin{aligned}b_{ijk}(x_{i},x_{j},x_{k};t+1) & =\psi_{ijk}(x_{i},x_{j},x_{k})e^{-1}e^{-\lambda_{ijk\rightarrow{i}}(x_{i})}\\
 & \times e^{-\lambda_{ijk\rightarrow{j}}(x_{j})}e^{-\lambda_{ijk\rightarrow{k}}(x_{k})}e^{-\gamma_{ijk}}
\end{aligned}
\end{alignat}

\begin{equation}
b_{i}(x_{i};t+1)=\psi_{i}(x_{i})e^{-1}e^{n_{i}}[b(x_{i};t)]^{n_{i}}
e^{\sum_{j}\sum_{k}\sum_{l}\lambda_{jkl\rightarrow{i}}(x_{i})-\alpha_{i}}
\end{equation}

Our inner loop equation

\begin{equation}
e^{\gamma_{ijk}(t+1)}=\sum_{x_{i},x_{j},x_{k}}\psi_{ijk}(x_{i},x_{j},x_{k})e^{-1}
e^{-\lambda_{ijk\rightarrow{i}}(x_{i};t)}e^{-\lambda_{ijk\rightarrow{j}}(x_{j};t)}
e^{-\lambda_{ijk\rightarrow{k}}(x_{k};t)}
\end{equation}
\begin{equation}
e^{2\lambda_{ijk\rightarrow{k}}(x_{k};t+1)}=\frac{\sum_{x_{i},x_{j}}\psi_{ijk}(x_{i},x_{j},x_{k})
e^{-\lambda_{ijk\rightarrow{i}}(x_{i};t)}e^{-\lambda_{ijk\rightarrow{j}}(x_{j};t)}
e^{-\gamma_{ijk}(t)}}{e^{n_{k}}(b_{k}(x_{k};t))^{n_{k}}e^{\sum_{o\neq i}{\sum_{p\neq j}\sum_{q\neq k}\lambda_{opq\rightarrow{k}}(x_{k};t)}-\alpha_{k}}}
\end{equation}
\begin{equation}
e^{2\lambda_{ijk\rightarrow{j}}(x_{j};t+1)}=\frac{\sum_{x_{k},x_{i}}\psi_{ijk}(x_{i},x_{j},x_{k})
e^{-\lambda_{ijk\rightarrow{k}}(x_{k};t)}e^{-\lambda_{ijk\rightarrow{i}}(x_{i};t)}
e^{-\gamma_{ijk}(t)}}{e^{n_{j}}(b_{j}(x_{j};t))^{n_{j}}e^{\sum_{o\neq i}{\sum_{p\neq j}\sum_{q\neq k}\lambda_{opq\rightarrow{j}}(x_{j};t)}-\alpha_{j}}}
\end{equation}
\begin{equation}
e^{2\lambda_{ijk\rightarrow{i}}(x_{i};t+1)}=\frac{\sum_{x_{j},x_{k}}\psi_{ijk}(x_{i},x_{j},x_{k})
e^{-\lambda_{ijk\rightarrow{j}}(x_{j};t)}e^{-\lambda_{ijk\rightarrow{k}}(x_{k};t)}
e^{-\gamma_{ijk}(t)}}{e^{n_{i}}(b_{i}(x_{i};t))^{n_{i}}e^{\sum_{o\neq i}{\sum_{p\neq j}\sum_{q\neq k}\lambda_{opq\rightarrow{i}}(x_{i};t)}-\alpha_{i}}}
\end{equation}

\begin{equation}
e^{\alpha_{i}}=e^{n_{i}-1}\sum_{x_{i}}[b_{i}(x_{i};t)]^{n_{i}}e^{\sum_{opq}\lambda_{opq\rightarrow{i}}(x_{i})}
\end{equation}

In this case we update$\lambda_{ijk}$lagrange multipliers by using
constraint condition and using normalization condition for the $\gamma_{ijk},\alpha_{i}$

Attention: we can update the marginal normalization lagrange multipliers
by the constraint condition $b_{i}(x_{i})=\sum_{x_{j},x_{k}}b_{ijk}(x_{i},x_{j},x_{k})$
when these beliefs guaranteed to satisfy the constraint
condition. In this case we can update the marginal normalization lagrange
multipliers by using the formula that indicated below
\begin{equation}
e^{\alpha_{i}}=\frac{e^{n_{i}}(b_{i}(x_{i};t))^{n_{i}}
e^{\sum_{opq}\lambda_{opq\rightarrow{i}}(x_{i};t)}}{\sum_{x_{j},x_{k}}\psi_{ijk}(x_{i},x_{j},x_{k})
e^{-\lambda_{ijk\rightarrow{i}}(x_{i};t)}e^{-\lambda_{ijk\rightarrow{j}}(x_{j};t)}
e^{-\lambda_{ijk\rightarrow{k}}(x_{k};t)}e^{-\gamma_{ijk}(t)}}
\end{equation}

Unfortunately the CCCP algorithm can't guaranteed to satisfy the constraint
condition, so we can't use this equation for updating.

\subsection{the form of the CCCP equations when zero temp}

When temperature equal to zero (or inverse temperature is infinite),
 the CCCP dynamics quickly gets trapped to a local minimal region of the
 Free energy landscape.

The outer loop equation

\begin{equation}
b_{ijk}(x_{i},x_{j},x_{k};t+1)=
 \begin{cases}
  0& \text{if $E_{\alpha}(x_{i},x_{j},x_{k})$=$1$},\\
  \psi_{ijk}(x_{i},x_{j},x_{k})e^{-1}\prod_{j\in\partial\alpha}
 e^{-\lambda_{j\rightarrow\alpha}(x_{j})}e^{-\gamma_{\alpha}}& \text{otherwise}
 \end{cases}
\end{equation}

\begin{equation}
b_{i}(x_{i};t+1)=\psi_{i}(x_{i})e^{-1}e^{n_{i}}[b(x_{i};t)]^{n_{i}}
e^{\sum_{opq}\lambda_{opq\rightarrow{i}}(x_{i})-\alpha_{i}}
\end{equation}

The inner loop equation

\begin{equation}
e^{\gamma_{ijk}(t+1)}=\sum_{x_{i},x_{j},x_{k},E_{\alpha}\neq1}\psi_{ijk}(x_{i},x_{j},x_{k})
e^{-1}e^{-\lambda_{ijk\rightarrow{k}}(x_{k};t)}e^{-\lambda_{ijk\rightarrow{i}}(x_{i};t)}
e^{-\lambda_{ijk\rightarrow{j}}(x_{j};t)}
\end{equation}
\begin{equation}
e^{2\lambda_{ijk\rightarrow{k}}(x_{k};t+1)}=\frac{\sum_{x_{i},x_{j},E_{\alpha}\neq1}\psi_{ijk}(x_{i},x_{j},x_{k})
e^{-\lambda_{ijk\rightarrow{i}}(x_{i};t)}e^{-\lambda_{ijk\rightarrow{j}}(x_{j};t)}
e^{-\gamma_{ijk}(t)}}{e^{n_{k}}(b_{k}(x_{k};t))^{n_{k}}e^{\sum_{o\neq i}\sum_{p\neq j}\sum_{q\neq k}\lambda_{opq\rightarrow{k}}(x_{k};t)-\alpha_{k}}}
\end{equation}
\begin{equation}
e^{2\lambda_{ijk\rightarrow{j}}(x_{j};t+1)}=\frac{\sum_{x_{k},x_{i},E_{\alpha}\neq1}\psi_{ijk}(x_{i},x_{j},x_{k})
e^{-\lambda_{ijk\rightarrow{k}}(x_{k};t)}e^{-\lambda_{ijk\rightarrow{i}}(x_{i};t)}e^{-\gamma_{ijk}(t)}}
{e^{n_{j}}(b_{j}(x_{j};t))^{n_{j}}e^{\sum_{o\neq i}\sum_{p\neq j}\sum_{q\neq k}\lambda_{opq\rightarrow{j}}(x_{j};t)-\alpha_{j}}}
\end{equation}

\begin{equation}
e^{2\lambda_{ijk\rightarrow{i}}(x_{i};t+1)}=\frac{\sum_{x_{j},x_{k},E_{\alpha}\neq1}\psi_{ijk}(x_{i},x_{j},x_{k})
e^{-\lambda_{ijk\rightarrow{j}}(x_{j};t)}e^{-\lambda_{ijk\rightarrow{k}}(x_{k};t)}
e^{-\gamma_{ijk}(t)}}{e^{n_{i}}(b_{i}(x_{i};t))^{n_{i}}e^{\sum_{o\neq i}\sum_{p\neq j}\sum_{q\neq k}\lambda_{opq\rightarrow{i}}(x_{i};t)-\alpha_{i}}}
\end{equation}

\begin{equation}
e^{\alpha_{i}}=e^{n_{i}-1}\sum_{x_{i}}[b_{i}(x_{i};t)]^{n_{i}}e^{\sum_{opq}\lambda_{opq\rightarrow{i}}(x_{i})}
\end{equation}

\end{widetext}

\section{implementation}

We implement the 3-sat CCCP algorithm in different constraint density
($\alpha$ respectively equal to 0.5 , 1.0 , 2.0 , 3.0 , 4.0 , 4.267),
we compared CCCP algorithm with BP algorithm in the circumstance that
BP algorithm converge, and give CCCP algorithm results alone in another
case. We selected the 10000 variables system and we use binary state
variables, each variable only have two states $x_{i}\in$\{-1 , +1\}.

Every clause connected with three variables randomly and the each
connection have two states $J_{a}^{i}\in$\{-1 , +1\}, the total clauses
number changes with constraint density $M=\alpha N$, where $M$ is
clauses number, $N$ is variables number, $\alpha$ is constraint
density. We randomly choose one map from clause configurations (the
total clause configurations number $\sigma(\underline{J})=2^{3M}$,
where $M$ is total clause number ) after the constraint density
be fixed, in our case there is no external field potentials.

We derived the Bethe free energy for the 3-sat problem and implemented
the CCCP and BP algorithm. We randomly initialized the lagrange multipliers,
we use unary marginal beliefs $b_{i}(x_{i})$ and unary joint beliefs
$b_{ijk}(x_{i},x_{j},x_{k})$. We used formula (58) for updating the
marginal normalization lagrange multiplier $\alpha_{i}$.

In the CCCP algorithm we used half parallel updating (parallel updating
in one clause but unparallel updating between clauses) rule for the
inner loop, it gives exact and stable results than parallel updating.
Although parallel updating converge very quickly, but it gives poor
results (marginal beliefs converge to 1 or 0, all the joint beliefs
whose energy equal to zero converge to identical , and the last one
joint belief whose energy equal to one converge to very small than
other beliefs).

In the CCCP algorithm we used one iteration for the inner loop and
one iteration for the outer loop in one step updating. This guaranteed
to converge very quickly than more inner loop iterations and still
gives good result. If we use more than one iteration for the inner
loop it leads the algorithm converges very slowly, even it does not
converge.

The CCCP algorithm guaranteed to satisfy the normalization condition
of marginal beliefs and joint beliefs, but it can't guaranteed to
satisfy the constraint condition, this is why we add new lagrange
multiplier $\alpha_{i}$ to guaranteed to convergence of the CCCP
algorithm.

In 2D and 3D spin glass (A.L.Yuille) the free energy and divergence
that getting by CCCP algorithm is vibrating in the first some iterations
(very small part of all iterations) and then monotonically decreasing
until to converge. So the free energy almost decreased monotonically.
But in our case the difference that getting by CCCP algorithm monotonically
decreasing until to converge. The free energy decreased monotonically
too when the constraint density smaller than 2.6, but it monotonically
decreasing in forepart and coming to increase in the back end (but
the increase very small almost no change) in other case, it is maybe
caused by the lost of data (it is related to property of computer
or we must simplify the updating formula) when programming. If we
ignore the lost of data, then we can say the free energy always converge
monotonically.

In the small system ($N<600$) we can get solution of 3-sat problem
by using CCCP algorithm, but in large system we can't.

In order to avoid lost of data when we add up two different data we
must avoid to add the biggest one to the smallest one as far as possible.

We implemented the BP algorithm in standard manner, we used a complete
parallel update rule.

BP algorithm can't guaranteed to decrease the free energy and divergence
of beliefs at each step of iterations. So it is more unstable than
CCCP algorithm. In our case the BP algorithm can't converge when constraint
density $\alpha_{i}$bigger than 3.86, but the CCCP algorithm still
converge in this case. The computation time very long along with constraint
density, but we can promote the computation speed by using lower convergence
precision (it gives almost same results with higher convergence precision,
if $\varepsilon_{1}=0.0001,\varepsilon_{2}=0.000001$ this two converge
precisions only leads 0.0001 and 0.000001 error respectively). The
beliefs that getting by BP algorithm don't satisfy the constraint
condition too until to converge.

BP algorithm and CCCP algorithm give same results when constraint
density smaller than 3.86 where BP algorithm can converge. In this
area we give simulation results only at $\alpha$=2.0 and $\alpha$=3.0
respectively, but in other circumstances our conclusion is still right.
We only give CCCP algorithm simulation results alone when constraint
density bigger than 3.86 where BP algorithm can't converge, and we
only give simulation results at $\alpha$=4.0 and $\alpha$=4.267(transition
point of 3-sat problem in 10000 points system) respectively, the CCCP algorithm still converge
when $\alpha>$4.267, but the result is not very meaningful.

\begin{figure}
\begin{centering}
\includegraphics[scale=0.35]{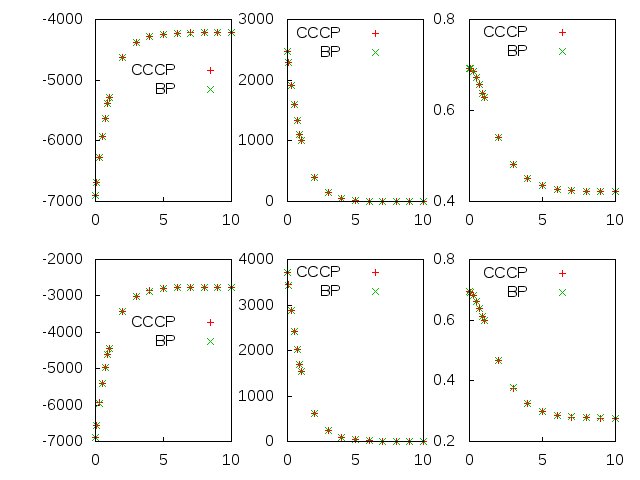}
\par\end{centering}

\caption{Compare results of CCCP algorithm and BP algorithm. Constraint density
$\alpha$=2.0 plots (Top Panels) and alfa=3.0 plots (Bottom Panels).
Left to Right the vertical axis represent Total Free Energy,Total
Mean Energy and Entropy Density respectively. The horizontal axis
represent the Inverse Temperature. The CCCP algorithm give same results
with BP algorithm in the BP algorithm converged area.}
\end{figure}

\begin{figure}
\begin{centering}
\includegraphics[scale=0.35]{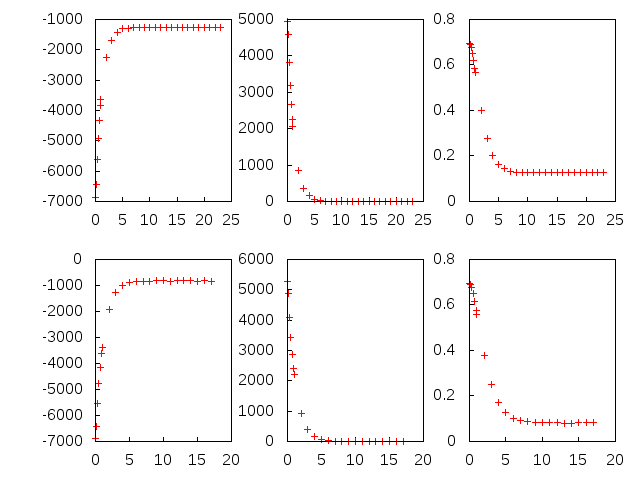}
\par\end{centering}

\caption{Performance of CCCP algorithm. Constraint density alfa=4.0 plots (Top
Panels) and alfa=4.267 plots (Bottom Panels). Left to Right the vertical
axis represent Total Free Energy,Total Mean Energy and Entropy Density
respectively. The horizontal axis represent the Inverse Temperature.
The CCCP algorithm still converge when Constraint Density bigger than
3.86, and the Mean Energy still converge to zero when the Inverse
Temperature bigger than some threshold value.}
\end{figure}

\begin{figure}
\begin{centering}
\includegraphics[scale=0.35]{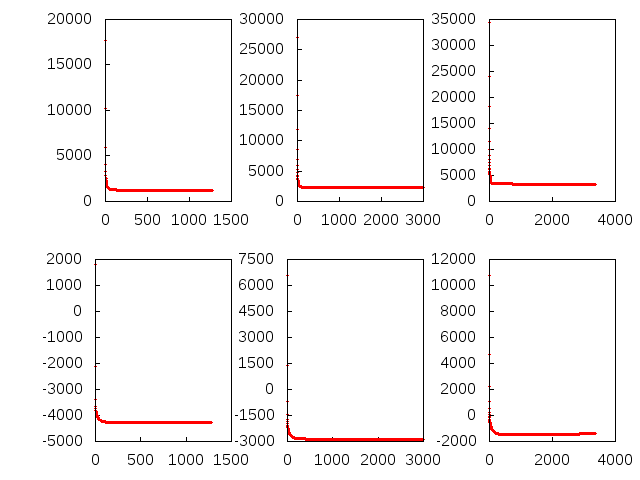}
\par\end{centering}

\caption{Performance of CCCP algorithm. Total Constraint difference plots (Top
Panels) and Total Free energy plots (Bottom Panels). Left
to Right the vertical axis represent The Constraint density Alfa=2.0,3.0,4.0
respectively. The horizontal axis represent the number of Iteration
Step. The constraint condition is not satisfied until to converge
for any constraint density, and the satisfiability of the constraint
condition decline with the increase of the constraint density.}
\end{figure}

\begin{figure}
\begin{centering}
\includegraphics[scale=0.35]{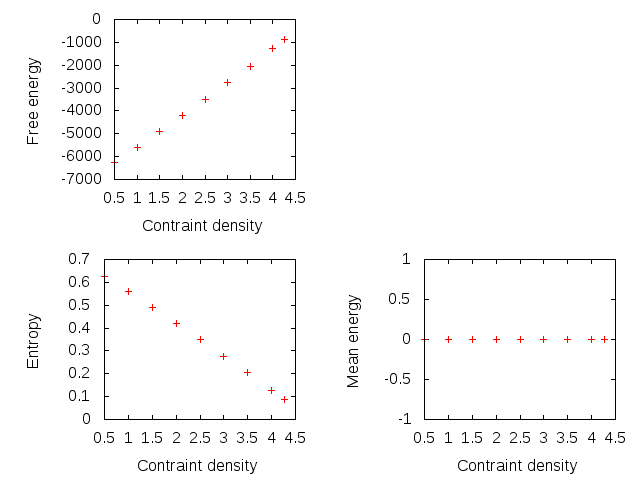}
\par\end{centering}

\caption{these are entropy, free energy and mean energy diagrams of CCCP algorithm
at zero temprature, the Y axis respectively represent the entropy,
free energy and mean energy, the X axis represent the constraint density
($\alpha=0.5,1.0,1.5,2.0,2.5,3.0,3.5,4.0,4.26$)}
\end{figure}

\section{Discussion}

It is known that BP algorithm is guaranteed to converge extrema point
of the Bethe free energy, although it gives exact solution on tree graphs,
but it amazingly successful on the graphs with no loops.
We can obtain the BP algorithm by using
Bethe approximation with statistical technic. Although It's converge
speed very quick, but it can't converge when constraint density bigger
than some threshold value. In our case the BP algorithm can't converge
when constraint density bigger than 3.86, this is a big disadvantages
of the BP algorithm .

The CCCP algorithm still guaranteed to converge to the extrema point
of the Bethe free energy when constraint density bigger than 3.86,
and it gives very similar results (the free energy, entropy, mean
energy) with the BP algorithm in the area where BP algorithm converge.
But these two algorithms have different changing rules on difference
and free energy in the whole convergence process. This property indicate
that although these two algorithm converge to same extrema points
of thermodynamical free energy functions, but the updating mechanism
totally different from each other. The BP algorithm update data which
only satisfy the normalization condition never consider about constraint
condition, but the CCCP algorithm update data which satisfy all the
condition theoretically (but it still doesn't satisfy the constraint condition
when programming, but it more and more close to satisfy the constraint condition
at each inner loop iteration step).

\section{conclusion}

In this paper we use CCCP principle to get double loop algorithm for
3-sat problem, this algorithm gives very same results with BP algorithm
where the BP algorithm can converge. The CCCP algorithm still converge
when BP algorithm can't converge.

The CCCP algorithm converge very stable than BP algorithm. In our
case the divergence monotonically decrease in all iteration process
for any constraint density. But the free energy monotonically decrease
in the forepart and begin to increase (but this increase very small
almost no changes this is related to how to set convergence precision)
in the back end when constraint density bigger than 2.6, actually
we can stop the iteration when free energy begin to increase and in
this time, we can still get accurate results, so we can say the free
energy (almost) always monotonically decrease.

Our CCCP double loop algorithm some different from Yuille's double
loop algorithm. We add marginal normalization lagrange multipliers
to the double loop algorithm, but he doesn't. If we don't add this
term to this algorithm, the double loop algorithm can't converge. It is because of we
can make up the error that leaded by inner loop equation(inner loop equation can't
guaranteed to satisfy the constraint condition and this directly affect
the beliefs of double loop algorithm, eventually this leads
NAN results of CCCP double loop algorithm) by adding this new variables.
 We use one iteration for inner
loop and updating data by half parallel way. If we use parallel updating
way, the CCCP algorithm can converge but gives wrong answer. Yuille
take five iterations and he not to say he updating the data by parallel
way or unparallel way. So our double loop algorithm has different
updating style from Yuille's double loop algorithm.
\begin{acknowledgments}
I would like thanks to Prof Haijun Zhou for helpful discussion and
KITPC(Beijing) for hospitality. This work was supported by the National Natural
Science Foundation of China with Grants No.......
(and the National Fundamental Research Program of China with Grants No...)
\end{acknowledgments}


\begin{thebibliography}{10}
\bibitem{Yuille2002} Yuille, A.L. Neural Computation, Vol. 14, No.
7, 1691-1722 (July 2002 )).

\bibitem{Hartmann2005}Hartmann, A.K. and Weigt, W., \char`\"{}Phase
Transitions in Combinatorial Optimization Problems\char`\"{}, Germany:
Wiley-VCH. (231-277 (2005)).

\bibitem{Mezard2009}Mezard,M. and A.Montanari, \char`\"{}Information,
Physics, and Computation\char`\"{}, Oxford University Press (197-217
(2009)).

\bibitem{Krzakala2007} Krzakala, F., Montanari, A., Ricci-Tersenghi,
F., Semerjian, G. and Zdeborova, L., \char`\"{}Gibbs states and the
set of solutions of random constraint satisfaction problems.\char`\"{},
PNAS 104 (25) (10318-10323 (2007)).

\bibitem{Monasson1996} Monasson, Rémi and Zecchina, Riccardo, \char`\"{}Entropy
of the K-Satisfiability Problem\char`\"{}, Phys. Rev. Lett. (1996),
3881--3885.

\bibitem{Domb1972} Domb, C. and Green, M.S., \char`\"{}Phase Transition
and Critical Phenomena\char`\"{}, Academic Press. London. (1972.).

\bibitem{Yedidia2000} Yedidia J.S., Freeman, W.T., Weiss, Y., \char`\"{}Bethe
free energy, Kikuchi approximation and belief propagation algorithms.\char`\"{},
NIPS 13 (689-695 (2000)).

\bibitem{Pearls1988} Pearls, J., \char`\"{}Probabilistic Reasoning
in Intelligent Systems\char`\"{}, Morgan Kaufman (177-184,241-250
(1988)).

\bibitem{Gallager1963} Gallager, R.G., \char`\"{}Low-density parity-check
codes.\char`\"{}, Combridge: MA: MIT Press (1963).

\bibitem{Freeman1999} Freeman, W.T. and Pasztor, E.C., \char`\"{}Learning
Low-Level Vision\char`\"{}, In.Proc. International Conference of Computer
Vision. ICCV'99 ((1999)), 1182-1189.

\bibitem{Frey1998} Frey, B.J, \char`\"{}Graphical models for machine
learning and digital communication\char`\"{}, MIT Press. (1998).

\bibitem{Murphy1999} Murphy, Kevin P. and Weiss, Yair and Jordan,
Michael I., \char`\"{}Loopy Belief Propagation for Approximate Inference:
An Empirical Study\char`\"{}, Morgan Kaufmann Publishers Inc. (1999),
467--475.

\bibitem{A.LYuille1995} A.L. Yuille, D. Geiger, \char`\"{}The Handbook
of Brain Theory and Neural Networks.\char`\"{}, MIT Press (1995).

\bibitem{Wang.c2013} Wang, C. and Zhou, H.-J., \char`\"{}Simplifying
generalized belief propagation on redundant region graphs\char`\"{},
Journal of Physics Conference Series (2013), 012004.

\bibitem{zhou2012} Zhou,H.J. and Wang,C., \char`\"{}Region Graph
Partition Function Expansion and Approximate Free Energy Landscapes:
Theory and Some Numerical Results\char`\"{}, Journal of Statistical
Physics (2012), 513-547.

\bibitem{Zhou2013} Jia Zeng and Cheung, W.K. and Jiming Liu, \char`\"{}Learning
Topic Models by Belief Propagation\char`\"{}, Pattern Analysis and
Machine Intelligence, IEEE Transactions on (2013), 1121-1134.

\bibitem{Zhou2011} Zhou H.J., Wang C., Xiao J.Q., Bi Z.D., \char`\"{}Partition
function expansion on region-graphs and message-passing equations\char`\"{},
Journal of Statistical Mechanics:Theory and Experiment (2011), L12001.

\bibitem{Shingh2013} Singh, S. and Riedel, S. and McCallum, A., \char`\"{}Anytime
Belief Propagation Using Sparse Domains\char`\"{}, ArXiv e-prints
(2013).

\bibitem{Ahamadi2013} Ahmadi, Babak and Kersting, Kristian and Mladenov,
Martin and Natarajan, Sriraam, \char`\"{}Exploiting symmetries for
scaling loopy belief propagation and relational training\char`\"{},
Machine Learning (2013), 91-132.

\bibitem{Gribova2002} Gribova, N.\textasciitilde{}V. and Tareyeva,
E.\textasciitilde{}E., \char`\"{}Replica Symmetry Breaking in an Axial
Model of Quadropolar Glass\char`\"{}, eprint arXiv:cond-mat/0204028
(2002).

\bibitem{Zeng2013} Zhou H.J., Zeng Y., \char`\"{}Solution Space Coupling
in the Random K-Satisfiability Problem\char`\"{}, Commun. Theor. Phys.
(2013).

\bibitem{Likang2009} Li, Kang and Ma, Hui and Zhou, Haijun, \char`\"{}From
one solution of a 3-satisfiability formula to a solution cluster:
Frozen variables and entropy\char`\"{}, Phys. Rev. E (2009), 031102.

\bibitem{Xiao2011} Xiao J.Q., Zhou H.J., \char`\"{}Partition function
loop series for a general graphical model: free energy corrections
and message-passing equations\char`\"{}, Journal of Physics A: Mathematical
and Theoretical (2011), 425001.

\bibitem{Crisanti2002} Crisanti, A. and Leuzzi, L., \char`\"{}First-Order
Phase Transition and Phase Coexistence in a Spin-Glass Model\char`\"{},
Phys. Rev. Lett. (2002), 237204.

\bibitem{Annibale2003} Annibale, Alessia and Cavagna, Andrea and
Giardina, Irene and Parisi, Giorgio, \char`\"{}Supersymmetric complexity
in the Sherrington-Kirkpatrick model\char`\"{}, Phys. Rev. E (2003),
061103.

\bibitem{J.Jkos1994} J.J. Kosowsky and A.L. Yuille, \char`\"{}The
invisible hand algorithm: Solving the assignment problem with statistical
physics\char`\"{}, Neural Networks (1994), 477 - 490.

\bibitem{Ranga1996} Rangarajan, A., Gold, S., Mjolsness, E., \char`\"{}A
Novel Optimization Network Architecture with Application.\char`\"{},
Neural Computation. (1996a.), 1041-1060.

\bibitem{Yuille1994} Yuille, A.L. and Kosowsky, J.J., \char`\"{}Statistical
Physics Algorithms that Converge.\char`\"{}, Neural Computation. (1994),
341-356.

\end{thebibliography}
\end{document}